# Mapping the Microscale Origins of MRI Contrast with Subcellular NV Diamond Magnetometry

Hunter C. Davis[1#], Pradeep Ramesh[2#], Aadyot Bhatnagar[3], Audrey Lee-Gosselin[1], John F. Barry[4,5,6,7], David R. Glenn[4,5,6], Ronald L. Walsworth[4,5,6], Mikhail G. Shapiro[1,*]

[1]Division of Chemistry and Chemical Engineering, [2]Division of Biology and Biological Engineering, [3]Division of Engineering and Applied Sciences, California Institute of Technology, Pasadena, CA 91125, USA

[4]Harvard-Smithsonian Center for Astrophysics, [5]Department of Physics, [6]Center for Brain Science, Harvard University, Cambridge, MA 02138, USA

[7]MIT Lincoln Laboratory, Lexington, MA 02420, USA

[#]Equal contribution

[*]Correspondence should be addressed to MGS: (mikhail@caltech.edu)

## Abstract

Magnetic resonance imaging (MRI) is a widely used biomedical imaging modality that derives much of its contrast from microscale magnetic field gradients in biological tissues. However, the connection between these sub-voxel field patterns and MRI contrast has not been studied experimentally. Here, we describe a new method to map subcellular magnetic fields in mammalian cells and tissues using nitrogen vacancy diamond magnetometry and connect these maps to voxel-scale MRI contrast, providing insights for in vivo imaging and contrast agent design.

## Main Text

Magnetic resonance imaging (MRI) is a common biomedical imaging modality whose contrast is based on the spatially and temporally varying behavior of nuclear spins. This behavior is influenced by microscale magnetic field gradients in cells and tissues produced by endogenous substances such as pathological iron deposits [1, 2] or molecular imaging agents such as iron oxide nanoparticles (IONs) [3, 4]. The resulting image contrast is used to diagnose diseases or track the *in vivo* distribution of labeled cells and molecules [5, 6]. The precise dependence of voxel-scale (~ 0.5 mm) MRI contrast on microscale magnetic field gradients has been a hot topic of theory and simulation due to its importance for disease diagnosis and contrast agent design [2, 7-10]. These studies predict, for example, that the spatial frequency of the local magnetic field can significantly impact the $T_2$ relaxation rate of a tissue and that optimizing contrast agent size can maximize $T_2$ contrast for a given set of material and imaging parameters. Despite the significance of the relationship between the spatial pattern of the local magnetic field and $T_2$ relaxation, this dependence has not been studied experimentally due to a lack of methods to map magnetic fields at the microscale under biologically relevant conditions.

Here, we present a method to study the connection between subcellular magnetic fields and MRI contrast using nitrogen vacancy (NV) magnetometry, a recently developed technique that enables the imaging of magnetic fields with optical resolution using the electronic properties of fluorescent NV quantum defects in diamond [11]. The electronic structure of an NV center forms a ground-state triplet, with the $m_s = \pm 1$ states separated from the $m_s = 0$ state by 2.87 GHz, making ground-state spin transitions addressable by standard electron spin resonance (ESR) techniques. The resonant frequency of these transitions shifts linearly in a magnetic field due to the Zeeman effect. Upon green laser excitation (532 nm), the $m_s = \pm 1$ states are more likely to undergo non-radiative relaxation than the zero-spin state, so that microwave-induced transitions from $m_s = 0$ to $m_s = \pm 1$ cause a drop in NV fluorescence. Thus, the local magnetic field of an NV center can be easily extracted from the optically reported ground-state spin transition frequency. Diamonds densely doped with NV centers make it is possible to optically image this resonant transition frequency over a wide field of view, thus providing an Abbe-limited image of the magnetic field at the diamond surface [12].

NV magnetometry has recently been used in proof-of-concept biological applications such as imaging the magnetic fields produced by magnetotactic bacteria [13], detecting magnetically labeled cancer cells [14], and measuring magnetic fields produced by neuronal action potentials [15]. However, to date this technology has not been used to map sub-cellular magnetic fields in living mammalian cells or to connect these maps to *in vivo* diagnostic imaging modalities such as MRI. Doing so requires adapting NV magnetometry for high-sensitivity imaging of sparse magnetic fields in cells and tissues, developing methods to convert 2-D NV data into the 3-D distribution of magnetic field sources and simulating the behavior of nuclear spins in the resulting magnetic fields. In addition, monitoring the evolution of magnetic fields in live cells requires operating under non-damaging optical and thermal conditions with reduced available signal. In this work, we address these challenges to enable the mapping of sub-cellular magnetic fields in an *in vitro* model of macrophage iron oxide endocytosis and an *in vivo* model of liver iron overload, connecting both to MRI contrast.

Our home-built NV magneto-microscope (**Fig. 1a**) was optimized for both high resolution magnetic field imaging of fixed samples and dynamic imaging of living cells. By virtue of a relatively thick NV layer in our diamond (~4 µm), we were able to significantly reduce the applied laser power compared to shallower surface-implanted NV diamond microscopes, while maintaining a strong NV fluorescent signal for rapid imaging. We used a total internal reflection geometry to minimize phototoxicity [13, 15] and bonded a silicon carbide wafer to the diamond base to improve thermal dissipation [15]. For cell imaging experiments, we applied a moderate bias field (~10 mT) to magnetize cell-internalized superparamagnetic IONs.

As a first test of our method, we imaged the magnetic fields resulting from the endocytosis of superparamagnetic IONs by murine RAW 264.7 macrophages. Magnetic labeling and *in vivo* imaging of macrophages is under development for a variety of diagnostic and therapeutic applications [4, 16-18], which could benefit from an improved understanding of the resulting MRI contrast. In particular, although labeling is typically done with dispersed particles, their internalization and subsequent compaction by the cell (**Fig. 1, b-c**) could produce radically different magnetic field profiles [8-10], which cannot be directly observed by conventional electron microscopy or iron staining techniques. We performed vector magnetometry on fixed macrophages after incubating them for one hour with 200 nm IONs and allowing one additional hour for internalization. After measuring the magnetic field along each of the four NV orientations (**Fig. 1d**), we projected the field maps along Cartesian axes convenient for magnetic dipole localization via orthogonalization and tensor rotation (**Fig. 1e**).

To convert the resulting 2-D vector map at the diamond surface into a 3-D model of magnetic fields inside the cell, we developed an algorithm for iterative localization of magnetic dipoles (**Fig. 1f**, **Supplementary Fig. S1**). First, the in-plane coordinates of putative dipole field sources (*e.g.*, clusters of magnetic particles) were identified from local minima in the x-component of the vector field, chosen parallel to the projection of the bias field onto the diamond surface. Then, the off-diamond height (z) and magnetic moment of each cluster were determined by fitting the local dipole field profile. After fitting the dipole at the strongest local minimum, the resulting magnetic field pattern was subtracted, and the next strongest local minimum fitted, with this process repeated until all local minima were exhausted. A global fit was then performed using the results from the local fits as starting parameters. The degree-of-freedom-adjusted $R^2$ for all the global fits made to 6 representative particle-containing cells was greater than 0.90.

To translate sub-cellular magnetic field maps into predictions about MRI contrast, we performed Monte Carlo simulations of nuclear spin $T_2$ decoherence in lattices of representative cells. These cells contained magnetic dipole distributions and magnitudes derived from NV magnetometry of a representative cellular library (**Fig. 2a**, **Supplementary Fig. S2**). This simulation predicted a bulk MRI $T_2$ relaxation time of 24.3 ms for a 1:1 mixture of supplemented and unsupplemented cells (**Fig. 2b**). When compared to an experimental MRI measurement of $T_2$ in macrophages prepared as in the NV experiment and pelleted in a 1:1 mixture with unsupplemented cells, the Monte Carlo prediction was accurate to within 2.8% (**Fig. 2, c-d**). Importantly, this relaxation time could not have been predicted solely from the concentration of IONs in the sample, as previous simulations have suggested a major influence of packing geometry on contrast agent relaxivity [8-10]. To establish that this relationship also holds for our model system, we performed MRI measurements and Monte Carlo simulations with IONs distributed in the extracellular space (**Fig. 2e**). Per iron mass, we found that this diffuse extracellular arrangement produces approximately 6.63-fold faster $T_2$ relaxation than do endocytosed particles, underlining the importance of microscale magnetic field patterns (**Fig. 2f**).

To extend this technique to diagnostic imaging, we performed NV magnetometry on liver specimens from a mouse model of hepatic iron overload. The spatial distribution of iron deposits in the liver and other tissues has been a topic of interest in clinical literature as an indicator of disease state, including efforts to discern it noninvasively using MRI [2]. Iron overload was generated through intravenous administration of 900 nm IONs to C57bl/6 mice (**Fig. 2g**). Livers were harvested 18 hours after injection and imaged with 7T MRI, showing enhanced macroscale $T_2$ relaxation compared to controls (**Fig. 2h**). To investigate the microscale nature of this contrast enhancement, we cryosectioned the livers of saline- and iron-injected mice and imaged the magnetic field profiles of these tissue sections on our NV magneto-microscope. We measured the projection of the magnetic field along a single NV orientation, probing the $m_s = 0$ to $m_s = +1$ and $m_s = 0$ to $m_s = -1$ transitions. This revealed a punctate distribution of magnetic dipoles within the liver tissue of the iron-overloaded mouse (**Fig. 2i**, **Supplementary Fig. S3**). We confirmed that these magnetic fields resulted from IONs using fluorescent imaging, for which purpose the IONs were labeled with a fluorescent dye. These results suggest that NV magnetometry could be used to map sub-voxel magnetic field patterns within clinical specimens.

Finally, we tested whether NV magnetometry could be used to follow the magnetic consequences of a dynamic process in living mammalian cells. Macrophages endocytosing IONs go through several stages of internalization, gradually reconfiguring diffuse particles into compacted lysosomal clusters (**Fig. 3a**). This process could be relevant to interpreting MRI data from labeled macrophages and to the development of clustering-based magnetic nanoparticle contrast agents [19, 20]. To image living cells, we adjusted our NV methodology to minimize optical and thermal energy deposition. We sub-sampled the NV spectrum to probe only the $m_s = 0$ to $m_s = +1$ transition of one NV orientation and limited laser illumination to 5 minutes per image. This allowed us to generate time-lapse images of magnetic fields coalescing inside macrophages after ION internalization (**Fig. 3, b-c**, **Supplementary Figs. S4-S5**). Cell viability (assessed via a Trypan Blue exclusion assay) was ~90%. To our knowledge, this represents the first magnetic field imaging of a dynamic process in living mammalian cells, and could aid the development of dynamic contrast agents for MRI.

In summary, this work has established methods to experimentally connect subcellular magnetic field gradients to voxel-scale MRI contrast and follow dynamic processes within living cells using NV diamond magnetometry. We anticipate that these methods will enable the microscale sources of MRI contrast to be studied in a variety of biological contexts to aid in the interpretation of clinical images and contrast agent development. While this study demonstrated the core capabilities of this method using iron-loaded cells and tissues, future improvements should extend this technique to samples containing weaker and more homogeneously distributed magnetic field sources. For example, employing diamonds with thinner NV layers would reduce the point spread function of NV-imaged magnetic fields, increasing the precision of source localization, while improved methods for positioning tissue sections flat on the diamond surface would allow the mapping of fields produced by smaller, endogenous magnetic inclusions. Combined with ongoing advances in NV diamond technology, these improvements will help subcellular magnetometry investigate the origins of MRI contrast in a broad range of biological and medical scenarios.

## Methods

*Nitrogen Vacancy Magneto-Microscope*
The NV magneto-microscope was constructed from a modified upright Olympus microscope and a 532 nm laser source. The diamond used in this work is an electronic grade (N < 5ppb) single crystal substrate with nominal rectangular dimensions of 4.5 mm x 4.5 mm x 500 μm, grown using chemical vapor deposition (CVD) by Element Six. The top-surface NV sensing layer is measured to be 3.87 μm thick, consists of 99.999% isotopically pure $^{12}C$ with 21.4 ppm $^{14}N$ ($3.77*10^{17}$ cm$^{-3}$) incorporated into the layer during growth. Layer thickness and nitrogen concentration were determined by secondary ion mass spectroscopy. The diamond was irradiated with an unknown irradiation dosage and then annealed first at 800 °C for twelve hours followed by 1000 °C for twelve hours. This diamond was affixed to a silicon carbide wafer (for enhanced heat dissipation), which was in turn affixed to a pair of triangular prisms to facilitate a total internal reflection excitation path. The prisms, silicon carbide wafer and diamond were fused using Norland Optical Adhesive (NOA 71). The diamond assembly was removable to allow live cell culture on the diamond surface in a cell culture incubator. Light was collected from the top of the diamond through a water-immersion objective. Images were acquired on a Basler acA2040-180kmNIR - CMV4000 CCD camera with 2048x2040 5.5 μm pixels (we used 256x1020 pixels to increase frame rate). For high-resolution vector magnetometry and tissue imaging, NV fluorescence was excited using a 100 mW Coherent OBIS LS 532 nm optically pumped semiconductor laser. For live cell imaging, we used an attenuated 2 W 532 nm laser from Changchun New Industries Optoelectronics. When necessary, focal drift was adjusted for using a piezo-driven stage (Thorlabs). Microwave pulses were applied through a single turn copper loop immediately surrounding the diamond. The microwave signal was generated by a Stanford Research Systems Inc. SG384 signal generator and amplified by a ZHL-16W-43-S+ amplifier from MiniCircuits. Experimental timing was controlled by a National Instruments USB 6363 X Series DAQ. A bias magnetic field was generated by two NeFeB grade N52 magnets (1"x2"x.5", K&J Magnetics) positioned on opposite sides of the NV diamond. The NV setup was controlled by custom software written in LabView.

*Cell Culture*
RAW 264.7 cells were cultured at 37º C and 5% $CO_2$ in DMEM (Corning Cellgro) and passaged at or before 70% confluence. For particle labeling, media was aspirated and replaced with phenol red-free DMEM supplemented with 279 ng/ml IONs (200 nm Super Mag Amine Beads Ocean Nanotech). After one hour, the ION solution was aspirated and cells were washed twice with PBS to remove unbound particles. For fixed-cell magnetometry, the cells were trypsinized quenched with DMEM and deposited on the diamond surface. After a 1 hour incubation at 5% $CO_2$ and 37 ºC, the cells were fixed with 4 % paraformaldehyde-zinc fixative (Electron Microscopy Services) and washed twice with PBS.

For live cell imaging, the cells were cultured as above until trypsinization and spotting on the diamond. Their media was supplemented with 0.1 mM ascorbic acid to mitigate phototoxicity [21]. For extended imaging (Supplementary Fig. S3a), the cells were maintained on the diamond in DMEM supplemented with 10 mM HEPES to stabilize pH at 7.4 under ambient atmosphere.

*Vector Magnetometry*
The bias magnetic field was aligned close to in-plane with the diamond surface while having sufficient out-of-plane field strength to resolve all NV resonances, and the full NV ODMR spectrum was probed. The microwave resonance for each pixel in the image was set as the center of the middle hyperfine peak of the transition. Spectra were swept at 0.5 Hz with 2000 images acquired per spectrum (0.9 ms exposure time). Images were acquired with an Olympus 60x water immersion objective (NA 1.0). Magnetometry spectra were acquired for 2 hours each. For a sub-set of measurements, this time was extended to 6 hours to improve SNR.

Projection field maps for each NV orientation were generated from the corresponding peaks in the NV ODMR, and a quadratic background subtraction was performed to remove the magnetic gradient from the bias magnets. Projection field maps were combined to form 3 orthogonal field maps with $B_z$ oriented normal to the diamond sensing surface. $B_x$ is defined as the projection of the applied field onto the diamond plane and $B_y$ is defined along the vector that completes the orthogonal set. Pixels were binned 2x2 in post-processing to boost SNR. (This does not cause a significant reduction in resolution as the binned pixels in the object plane are 92 nm on a side, which oversamples the Abbe limit of ~340 nm.) Our sensitivity for this technique was 17 nT at 1 µm in plane resolution (our out of plane resolution was fixed by the NV layer depth of 3.87 µm). Of particular note, this detection limit is sufficient to detect a single 200 nm ION spaced 10 µm off of the diamond, where the peak measured $|B_x|$ field would be greater than 200 nT.

*Live Cell Magnetometry*
For live cells, the bias magnetic field was aligned along an arbitrary direction sufficient to resolve at least one NV resonance, and the magnetic field projection along a single NV orientation was probed using the $m_s = 0 \rightarrow m_s = +1$ transition. The microwave resonance for each pixel in the image was set as the center of the middle hyperfine peak of the transition. Spectra were swept ~10 MHz at 1 Hz with 200 images acquired per spectrum (4 ms exposure time). In order to limit phototoxicity, each image was averaged for only 5 minutes and the laser was shuttered for five minutes in between images, resulting in a 50% duty cycle. Regions of interest were selected to include all relevant fields for a given cell. Optical power density was ~40 W/cm². Images were acquired with a Zeiss 40x near infrared water immersion objective (NA 0.8). Cell viability was assessed by performing a Trypan Blue exclusion assay at the conclusion of NV measurements.

*Field Fitting and Dipole Localization*
In-plane dipole coordinates were identified as local minima in the $B_x$ field map. Before localization, the field map was spatially low-passed (2D Gaussian filter with $\sigma = 0.5$ pixels) to eliminate noise-generated local minima in the background. A pixel was identified as a local minimum if and only if its $B_x$ field value was smaller than all of its immediate neighbors (including diagonals) in the spatially low-passed image.

Starting with the strongest local minimum, the measured magnetic field in a 10x10 pixel (~1.8x1.8 µm) square surrounding this minimum was fitted to a point dipole equation and averaged through the full NV layer depth (assuming uniform NV density), with the magnetic moment, height off of the diamond, and dipole orientation as free parameters.

$$B_x(i,j) = \frac{\int_{-z}^{-(z+h)} B_{xo}(i',j',b,M,\theta,\phi) \cdot db}{-h}$$

where

$$B_{xo}(i,j) = \frac{\mu_0}{4\pi} \cdot \left( \frac{3\,x\,(\boldsymbol{M} \cdot \boldsymbol{r})}{r^5} - \frac{\boldsymbol{M} \cdot \hat{x}}{r^3} \right)$$

Here i'=(i-i$_0$) and j'=(j-j$_0$) where (i$_0$,j$_0$) are the in-plane coordinates of the magnetic dipole, $\theta$ and $\phi$ correspond to the two rotational degrees of freedom available to a point dipole, M is the magnetic moment, z is the height of the dipole over the diamond, **r** is the displacement vector, $\hat{x}$ is the unit vector along the projection of the dipole axis onto the diamond surface plane, $x = i' \cos(\theta) - j' \sin(\theta)$, b is a dummy variable for integration through the NV layer, and h is the NV layer thickness. All parameters are free to fit other than the in-plane dipole coordinates, which are fixed by the local minimum of the $B_x$ field map.

After the strongest minimum has been fitted, the fitted field from the fit dipole (within the full field of view) was subtracted from the magnetic field image, to facilitate the fitting of weaker dipoles. The 10x10 pixel neighborhood of the second strongest dipole was then fitted in the subtracted image. The fitted field was subtracted, and the fitting continued until the list of local minima had been exhausted.

A global fit was then performed using the results from the neighborhood fits as starting parameters. The global fit function is the sum of N dipoles (where N is the number of local minima) with the in-plane dipole coordinates fixed at the local minima.

$$B_{x_{tot}}(i,j) = \sum_q B_{x_q}(i,j)$$

Here q is an index that runs from one to N and indicates the dipole field source. The precision of this technique is limited by the key assumption that the local minima are not significantly shifted in the x-y plane by neighboring dipoles. The degree of freedom-adjusted $R^2$ for each of the four global fits in the cell library was greater than 0.9. (For 3 of the 6 labeled cells, with image acquisition time increased from 2 to 6 hours, the $R^2$ was greater than 0.95) While this approach was able to produce a sufficiently precise magnetic field reconstruction to predict MRI relaxation, other methods are also available for analytic dipole localization and magnetic field reconstruction [22].

*Monte Carlo Simulations & Cell Library*
Nuclear spin relaxation was simulated by assigning 11 representative cells from vector magnetometry to random positions in a repeating face-centered cubic lattice containing a total of 108 spherical cells with periodic boundary conditions. Cell size was set to match previously measured values for RAW 264.7 macrophages [23]. Water molecules were randomly assigned initial x, y, and z coordinates in the lattice and allowed to diffuse while their phase in the rotating frame evolved from $\phi(0) = 0$ by $\delta\phi(t) = -\gamma B(x,y,z)\delta t$, where $B(x,y,z)$ is the local nanoparticle-induced field. Re-focusing pulses were simulated at 5.5 ms Carr-Purcell time by setting $\phi(t) = -\phi(t - \delta t)$. The magnetic field was mapped within this 3D-volume using a finite mesh whose mesh size was inversely proportional to the local field gradient. If a water molecule moved within a distance equivalent to six nanoparticle cluster radii of a cluster, the field contribution from that cluster was calculated explicitly. Background RAW cell relaxation was accounted for by post-multiplying the simulated signal with an exponential decay with time constant set to the measured relaxation rate of an unlabeled RAW cell pellet. Cell membranes were modeled as semi-permeable boundaries with a permeability of .01 $\frac{\mu m}{ms}$ in accordance with previously measured values for murine macrophage-like cells, adjusted to the temperature in our magnet bore (12.9 ºC) [24]. Intracellular and extracellular water diffusivity were set, respectively, to 0.5547 and 1.6642 $\frac{\mu m^2}{ms}$ in accordance with previously established values for water diffusivity at 12.9 ºC [25]. Bulk spin magnetization in the sample was calculated as $M(t) = \sum_i \cos[\phi_i(t)]$, where *i* is the index of simulated water molecules. Nanoparticle clusters were modeled as spheres packed so as to occupy three times the volume of their constituent nanoparticles, consistent with other simulations and grain packing theory [10] [26]. To account for the increase in nanoparticle magnetizations at 7T compared to our NV bias field, we scaled dipole magnetization using a SQUID-measured curve (Supplementary Fig. S6). Data presented in the manuscript represents the output of N=10 simulations, each containing 20 random arrangements of cells and 2000 water molecules.

To assess the impact of an alternative nanoparticle distribution (Fig. 2, e-f), we simulated the same quantity of nanoparticles, unclustered and distributed randomly in the extracellular space. The presented data comprises N=10 simulations, each containing 20 random arrangements of particles and 2000 water molecules.

*MR Imaging and Relaxometry*
Imaging and relaxometry were performed on a Bruker 7T MRI scanner. A 72 mm diameter volume coil was used to both transmit and receive RF signals. To measure the T$_2$ relaxation rate of RAW cells after nanoparticle labeling, the cells were labeled identically to their preparation for NV magnetometry, then trypsonized, quenched with 10 mL DMEM and pelleted for 5 min at 350 g. DMEM was aspirated and cells were resuspended in 150 µL PBS. After transferring the cells to a 300 µL centrifuge tube, they were pelleted for 5 min at 350 g. These tubes were embedded in a phantom comprising 1% agarose dissolved in PBS and imaged using a multi-echo spin echo sequence (TR = 4000 ms, TE = 11 ms, 2 averages, 20 echoes, 273x273x1000 µm voxel size). $T_2$ relaxation was obtained from a monoexponential fit of the first 6 echoes. (The slower background relaxation rates of four unsupplemented RAW cell pellets were measured using the first 20 echoes.)

For the scenario in which nanoparticles are unclustered in the extracellular space, unsupplemented RAW cells were pelleted and re-suspended in PBS supplemented with 100 µg/ml IONs. This concentration was selected to ensure a measurable $T_2$ and allow both *in silico* and *in cellulo* comparisons between the per-iron relaxation rates of extracellular and internalized particle scenarios. The validity of a per-iron comparison was confirmed by previous studies of the linearity of relaxivity for this size of iron oxide nanoparticles [27]. After supplementation, the cells were re-pelleted and immediately imaged as described above.

*Mouse Model of Iron Overload*
C57bl/6 mice were injected in the tail vein with 10 mg/kg of dragon green labeled 900 nm ION (Bangs) or saline. 16 hours after injection, the mice were perfused with 2 mL of 10% formalin, and their livers were harvested for MRI or NV magnetometry. MRI was performed on livers embedded in 1 % agarose using the 7T scanner described above, using a spin-echo pulse sequence with TR = 2500, TE = 11 ms, 4 averages, and a 273x273x1000 µm voxel size. For NV magnetometry, the liver was frozen in OCT embedding media and sectioned into 10 µm slices. Sections were mounted in on glass coverslips. We inverted the glass cover slip and pressed the tissue sample against the NV diamond. Silicon vacuum grease was applied at the edge of the cover slip (away from the diamond) to seal the sample against the diamond. After this preparation was complete, PBS was added to the dish to wet the sample. To compensate for magnetic field sources being further from the diamond due to tissue thickness and/or folds in the sections, NV imaging was performed with a strong (~250 G) bias field applied along a single NV axis. This strong bias field served to increase the magnetization of the magnetic inclusions in the liver. As it was applied along an NV axis, this bias field did not significantly reduce the contrast of the relevant ODMR spectral lines. Images were acquired with a Zeiss 40x near infrared water immersion objective (NA 0.8).

*Software and Image Processing*
All fits and plots were generated in MATLAB. Monte Carlo Simulations were performed in C++ on a Linux high performance computing cluster.

# Acknowledgements

We acknowledge Arnab Mukherjee, George Lu, My Linh Pham, Andrei Faraon, Geoffrey Blake, Joe Kirschvink, Hans Gruber, Michael Tyszka, Russ Jacobs, and John Wood for helpful discussions. This work was supported by the National Science Foundation Graduate Research Fellowship (PR), Caltech Center for Environmental - Microbial Interactions (MGS), the Burroughs Wellcome Fund (MGS), the NSF EPMD and PoLS programs (RLW) and the ARO MURI program on Imaging and Control of Biological Transduction using NV-Diamond (RLW). Research in the Shapiro Laboratory is also supported by the Heritage Medical Research Institute.

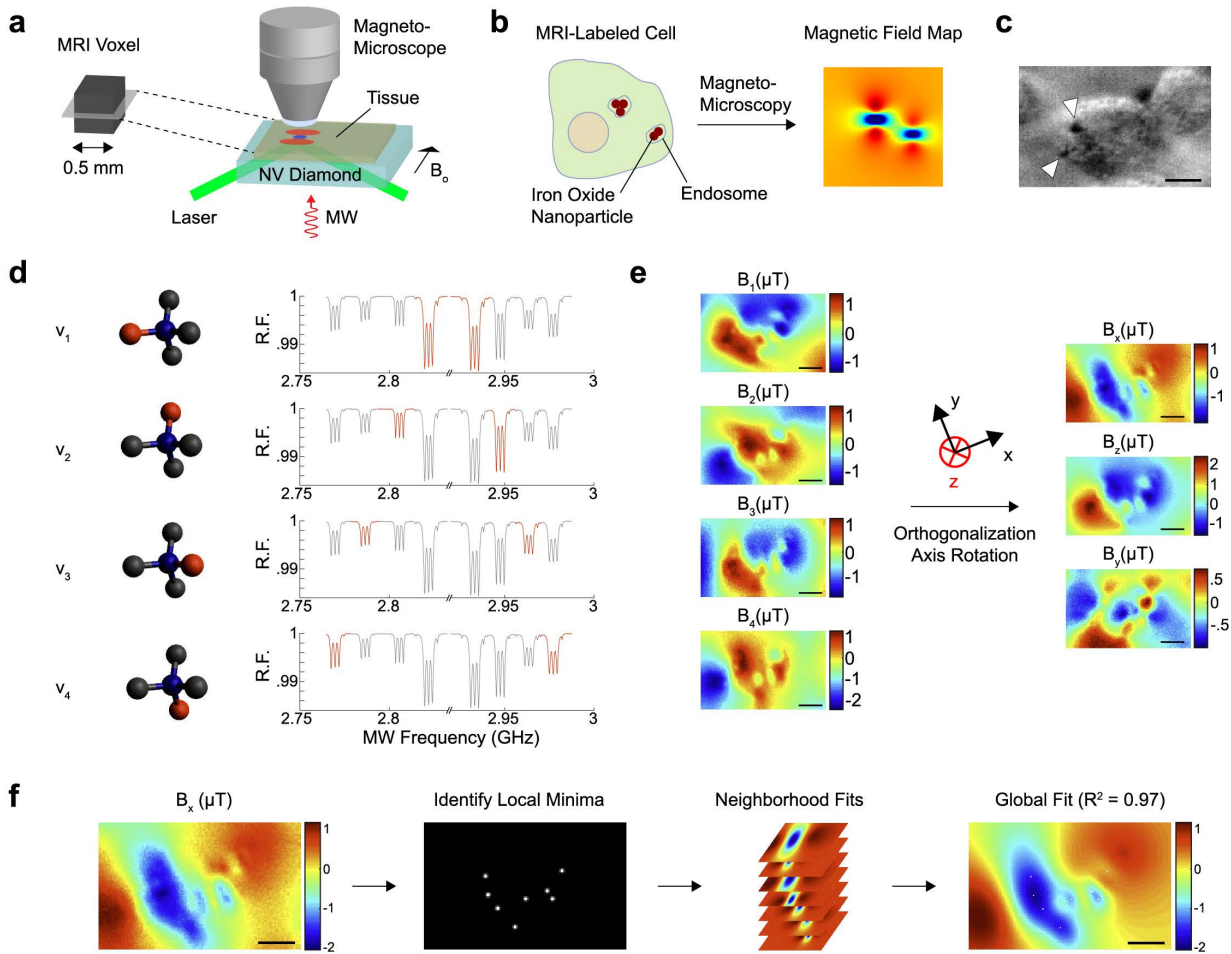

**Figure 1 – Sub-cellular mapping of magnetic fields in cells labeled for MRI.** (**a**) Schematic of sub-voxel magnetic field mapping using a NV magneto-microscope. (**b**) Illustration of a cell labeled with IONs and its expected magnetic field pattern. (**c**) Brightfield image of RAW 264.7 macrophage labeled with 200 nm IONS. White arrows point to internalized IONs. A brightfield imaging artifact also appears as black in the upper right corner of the cell. (**d**) Cartoon representation of each NV orientation and corresponding representative spectra from fixed cell experiments. The blue ball represents the nitrogen and the red ball represents the adjacent lattice vacancy. Highlighted peaks in each relative fluorescence (RF) spectrum show the transition corresponding to each of the 4 orientations. (**e**) Magnetic field images of the field projections along each of the 4 NV axes of macrophages 2 hours after initial exposure to 279 ng/ml 200 nm IONs (left). These images are converted via Gram-Schmidt orthogonalization and tensor rotation to field maps along 3 Cartesian coordinates with the z-axis defined perpendicular to the diamond surface and the x-axis defined as the projection of the applied bias field onto the diamond surface plane (right). The y-axis is defined to complete the orthogonal basis set. (**f**) Representative example of the procedure for dipole localization in cellular specimens. This procedure comprises three steps: first the local minima in the field map are identified and ranked; next, in decreasing order of magnitude, the neighborhood of each local minimum is fit to a point dipole equation and the resulting field is subtracted from the field map to reduce the fit-deleterious effect of overlapping dipole fields; finally, the results of these fits are used as guess parameters for a global fit over the full field-of-view. The fit shown has a degree-of-freedom-adjusted $R^2$ of 0.97. Scale bars are 5 μm.

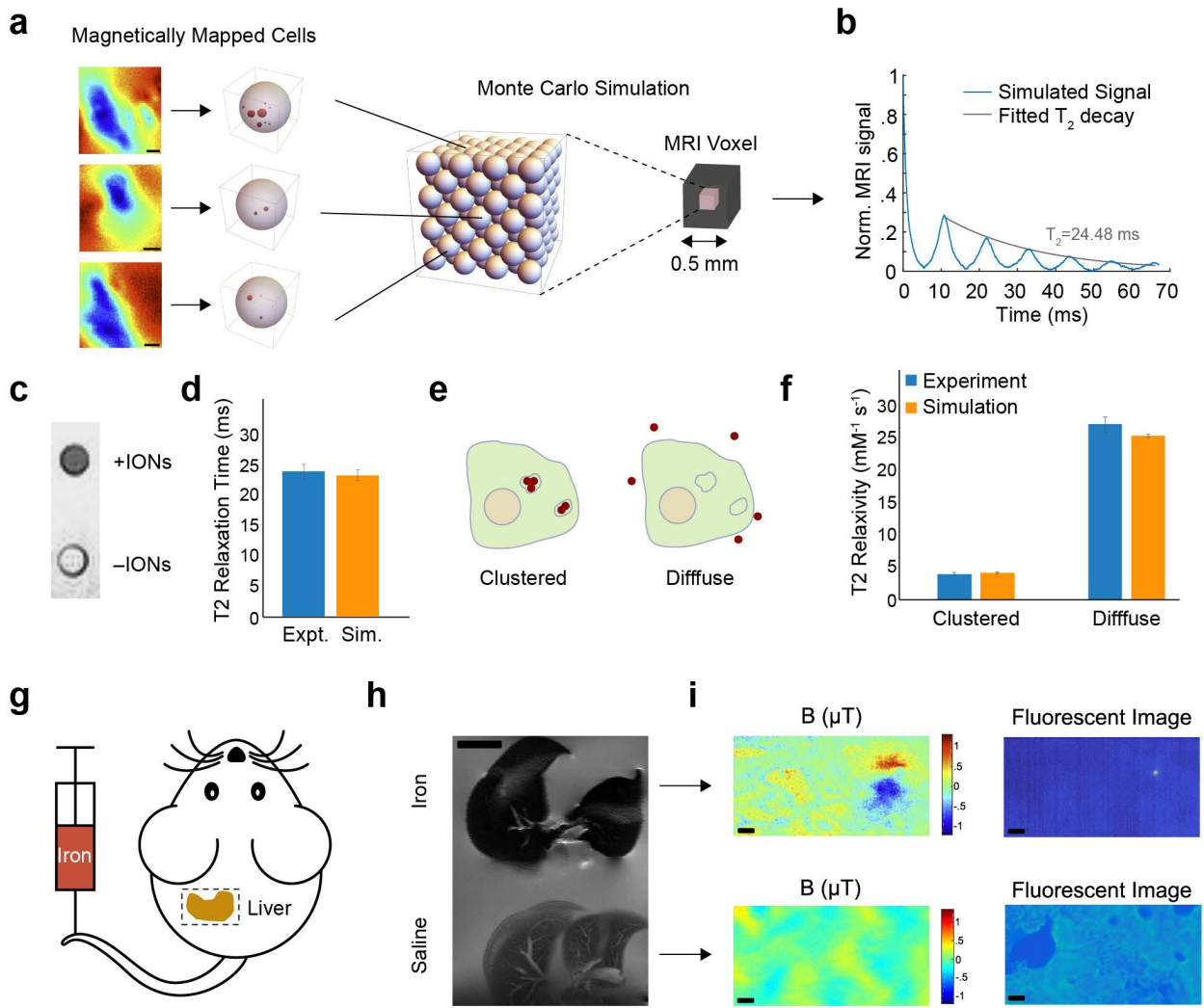

**Figure 2 – Predicted and experimental MRI behavior in cells and tissues.** (**a**) Schematic of Monte Carlo modeling of spin relaxation using NV-mapped magnetic fields. A library of 11 cells mapped with vector magnetometry (three representative cells shown) in a 1:1 mix with unlabeled cells, was used to randomly fill a 108 cell FCC lattice with periodic boundary conditions and run a Monte Carlo simulation of spin echo MRI to predict $T_2$ relaxation behavior. (**b**) Representative simulated MRI signal. (**c**) $T_2$-weighted MRI image of cell pellets containing a 1:1 mixture of supplemented and unsupplemented cells (top) or 100% unlabeled cells (bottom). (**d**) Simulated and experimentally measured $T_2$ relaxation times for the 1:1 mixture. (**e**) Illustration of the same quantity of magnetic particles endocytosed or distributed in the extracellular space. (**f**) Simulated and experimentally measured relaxivity for endocytosed and extracellular distributions of IONs. (**g**) Diagram of mouse model of iron overload, prepared by injecting 10 mg/kg of 900 nm iron oxide nanoparticles into the tail vein. (**h**) 7T $T_2$-weighted MR image of fixed, excised mouse livers from mice injected with IONs or saline. (**i**) NV magnetic field maps and fluorescence images of 10 μm liver sections obtained from the mice in (h). (Scale bars in a, h, and i are 2.5 μm, 5 mm, and 10 μm respectively.) Measurements and simulations have N=5 replicates. Error bars represent S.E.M.

Davis, Ramesh et al. Page 8 of 14

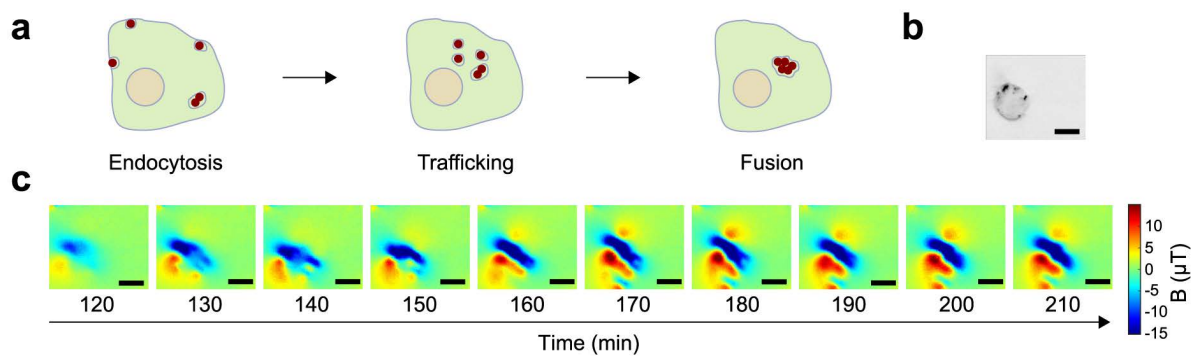

**Figure 3 – Dynamic magnetic microscopy in live mammalian cells.** (**a**). Cartoon showing the typical progression of endocytotic uptake of IONs. (**b**) Brighfield image of RAW macrophage used for live cell imaging in (c). (**c**) Series of time lapse magnetic field images of an individual macrophage endocytosing IONs. Scale bars are 10 µm.

# Supplementary Information

*Verification that a Live Cell Process was required for nanoparticle trafficking*

To ensure that the nanoparticle motion observed in the live cell experiments was due to a love cellular process, we also analyzed particle motion in fixed cells. Figure S5d shows that there was no observable change in the nanoparticle fields over the relevant time-course in fixed cells (~10 hours).

*SQUID Magnetometry*

To enable the use of NV magnetometry results to predict MRI behavior in Monte Carlo simulations we scaled the the measured magnetic moments from the 10 mT NV bias field to the 7T MRI field strength using the results of SQUID magnetometry performed on the relevant IONs (Figure S6). As our 7T MRI far exceeds the necessary bias field to saturate the nanoparticles, we assume that the nanoparticles are all at their saturation magnetization and their magnetic moments are all aligned along the bias field.

**Supplementary Figures**

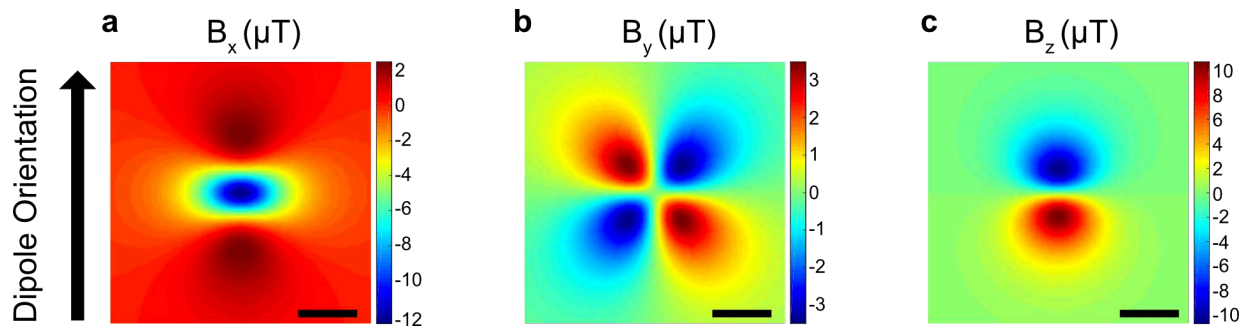

**S1 - Simulated dipole fields.** Simulated $B_x$ (**a**), $B_y$ (**b**), and $B_z$ (**c**) field projections for a point dipole oriented towards the top of the image with a magnetic moment of $10^{-15} A \cdot m^2$. The x and y coordinates of the dipole are fixed at the center of the image and the dipole is spaced two µm above the plane of projection. As in the main text, x is defined along the dipole axis, z is defined out of the page, and y is defined to complete the normal basis. Scale bars are 2.5 µm.

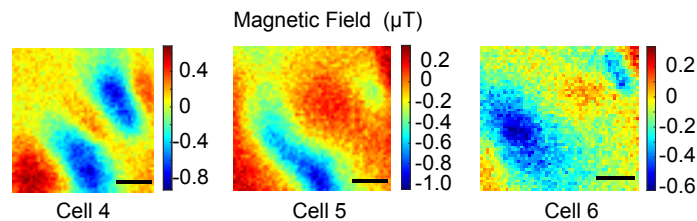

**S2 - Additional cells for Monte Carlo library.** Vector magnetometry results from three additional cells. These cells were measured as described in Figure 1 with the exception that the imaging time was cut to 2 hours. Scale bars are 2.5 µm.

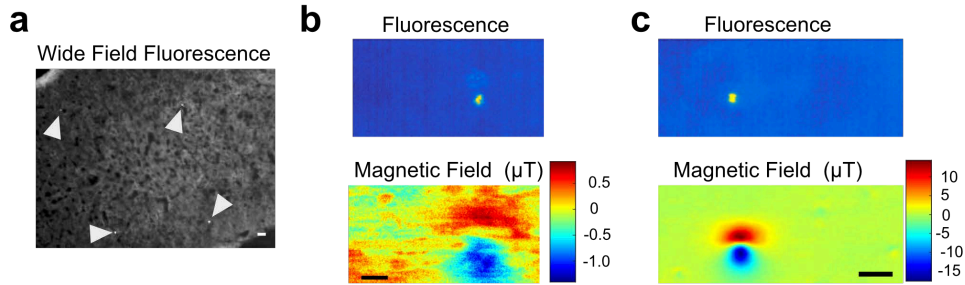

**S3 - Additional tissue sections.** (**a**) Fluorescent image of a wide field of view of a representative liver tissue section from an iron-injected mouse. Punctate fluorescent spots from the fluorescently labeled 900 nm ION are sparsely visible in the fluorescent image. (**b-c**) Field profile of two additional clusters measured using our NV microscope, measured as in Figure 2. Scale bars are 20 µm.

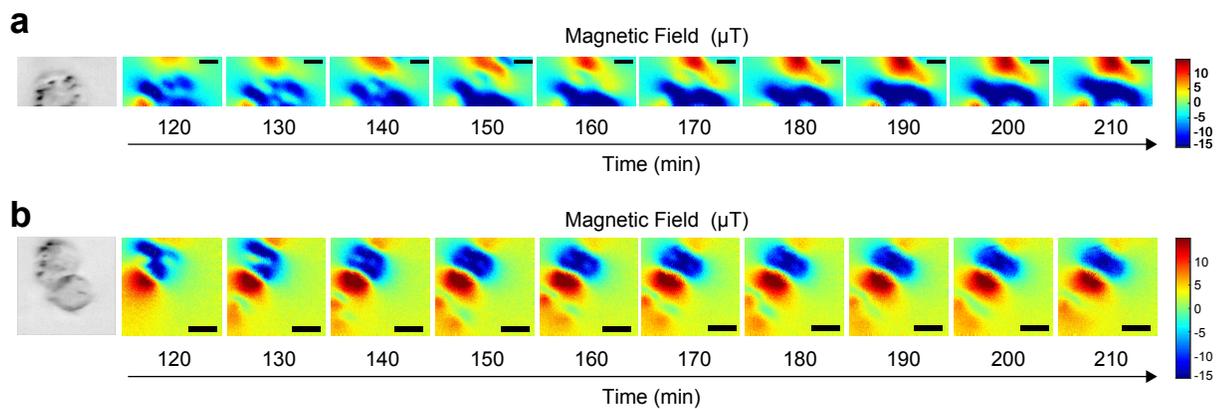

**S4 − Additional live cells.** (**a-b**) Two additional live cell replicates matching Figure 3b. Cells were confirmed alive with trypan blue after NV imaging. Scale bars are 5 µm.

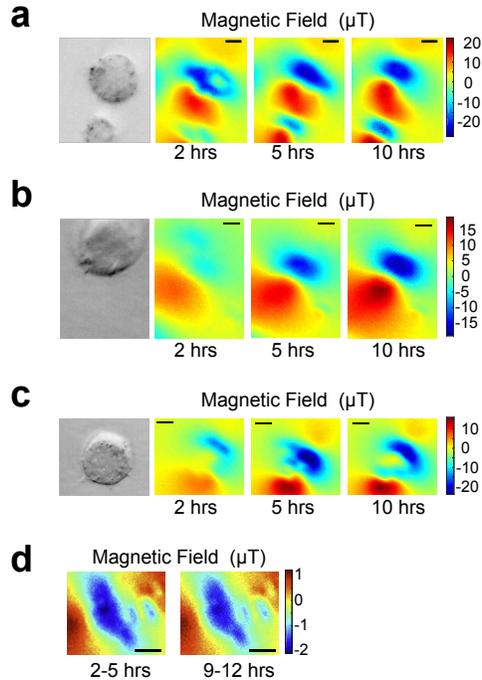

**S5 - Live cell imaging with extended time course.** (**a-c**) Live cells images of ION endocytosis acquired 2, 5, and 10 hours after initial nanoparticle exposure to 279 ng/ml 200 nm IONs. Trypan blue assay revealed an ~70% viability for these imaging studies. All displayed cells were still alive after imaging. (**d**) Magnetic field map from fixed cell acquired 7 hours apart to show the absence of dynamic changes in the magnetic field. Scale bars are 5 μm.

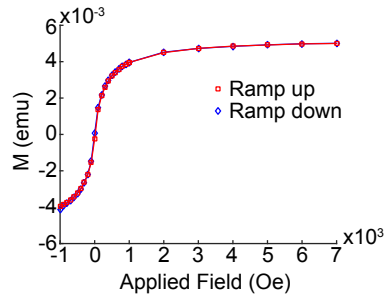

**S6 - SQUID magnetometry of IONs.** To measure the field dependent magnetic susceptibility of our ION sample, we performed SQUID magnetometry on a 100 ug stock at room temperature.